\documentclass[english,preprint]{elsarticle}
\usepackage[T1]{fontenc}
\usepackage[latin9]{inputenc}
\usepackage{esint}

\usepackage{babel}

\newcommand{\R} {\mbox{Re}\,}
\newcommand{\I} {\mbox{Im}\,}

\newcommand{\ee}{\end{equation}}
\newcommand{\be}{\begin{equation}}

\newcommand{\bea}{\begin{eqnarray}}
\newcommand{\eea}{\end{eqnarray}}
\newcommand{\ba}{\begin{eqnarray}}
\newcommand{\ea}{\end{eqnarray}}
\newcommand{\eu}{{\rm e}}
\newcommand{\ii}{{\rm i}}

\newcommand{\de}{{\displaystyle\rm\mathstrut d}}
\newcommand{\Ord}{{\rm O}}

\def\XXint#1#2#3{{\setbox0=\hbox{$#1{#2#3}{\int}$}
     \vcenter{\hbox{$#2#3$}}\kern-.5\wd0}}

\begin{document}

\title{Emptiness and Depletion Formation Probability in spin models with inverse square interaction}

\author[ICTP]{Fabio Franchini}
\ead{fabio@ictp.it}
\author[stonybrook,brookhaven]{Manas Kulkarni}
\ead{kulkarni@grad.physics.sunysb.edu}

\address[ICTP]{The Abdus Salam ICTP; Strada Costiera 11, Trieste, 34100,
Italy}
\address[stonybrook]{Department of Physics and Astronomy, Stony Brook University, Stony Brook, NY 11794-3800}
\address[brookhaven]{Department of Condensed Matter Physics and Materials Science, Brookhaven National Laboratory, Upton, NY-11973}

\begin{abstract}
We calculate the Emptiness Formation Probability (EFP) in the spin-Calogero Model (sCM) and Haldane-Shastry Model (HSM) using their hydrodynamic description. The EFP is the probability that a region of space is completely void of particles in the ground state of a quantum many body system. We calculate this probability in an instanton approach, by considering the more general problem of an arbitrary depletion of particles (DFP). In the limit of large size of depletion region the probability is dominated by a classical configuration in imaginary time that satisfies a set of boundary conditions and the action calculated on such solution gives the EFP/DFP with exponential accuracy. We show that the calculation for sCM can be elegantly performed by representing the gradientless hydrodynamics of spin particles as a sum of two spin-less Calogero collective field theories in auxiliary variables. Interestingly, the result we find for the EFP can be casted in a form reminiscing of spin-charge separation, which should be violated for a non-linear effect such as this. We also highlight the connections between sCM, HSM and $\lambda=2$ spin-less Calogero model from a EFP/DFP perspective.
\end{abstract}

\begin{keyword}
EFP \sep DFP\sep Integrable Models\sep Calogero-Sutherland \sep Haldane-Shastry \sep Hydrodynamics
\sep Collective Field Theory

\PACS 71.10.Pm,  75.10.Pq, 02.30.Ik,  03.75.Kk,  71.27.+a
\end{keyword}

\maketitle

\tableofcontents

\section{Introduction}

One-dimensional integrable models have an important role in the study of strongly correlated systems. When the reduced dimensionality makes interaction unavoidable, perturbative techniques can quickly loose applicability and over the years more sophisticated tools have been developed to tackle these problems. These tools clearly involve certain approximations and the existence of an exact solution for some models can allow to check their validity.

The conventional approach in solving quantum integrable model is known as Bethe Ansatz (and its generalization). It is very successful in constructing the thermodynamics of a system, but not very suitable to study its dynamics and the correlation functions, due to the increasing complexity of its solutions. However, a very elegant formalism was developed using the Quantum Inverse Scattering Method (QISM) \cite{QISM} to express correlation functions as determinants of certain integral operators (Fredholm determinants). In this formalism, the simplest correlation function one can write is known as the {\it Emptiness Formation Probability} (EFP) and measures the probability $P(R)$ that a region of length $2R$ is completely void of particles. For lattice models, one is interested in $P(n)$, the probability that $n$ consecutive lattice sites are empty. In spin chain, taking advantage of the Jordan-Wigner mapping between particles and spins, the same quantity can be thought of as the {\it Probability of Formation of Ferromagnetic Strings} (PFFS), i.e. the probability that $n$ consecutive spins are aligned in the same direction.

One should notice that the EFP is an $n$-point correlator and is, therefore, a much more complicated object compared to the usual two-point correlation functions one normally studies in condensed matter physics. However, due to the strongly interacting nature of the 1-D model, the QISM tells us that it is in fact no worse than other correlators between two points a length $n$ apart and even somewhat simpler and more natural. Moreover, the EFP is one of those extended objects like the Von Neumann Entropy, or the Renyi Entropy, that in recent years have attracted a lot of interest because of their ability to capture global properties that were not observed before from the study of $2$-point correlation functions. The latter quantities are of course motivated by studies of entanglement and quantum computation, while the EFP arises naturally in the contest of integrable theories.

Despite the claimed simplicity, the calculation of the EFP is by no means an easy task. For some models, the specific structure of the solution has allowed to find the asymptotic behavior of the EFP as $n \to \infty$. For instance, the EFP in the whole of the phase-diagram of the $XY$ model was calculated in \cite{shiroishi01, abanovfran03, franabanov05, Fabio-Thesis} using the theory of Toeplitz determinants, while for the critical phase of the $XXZ$ spin chain the solution was found in \cite{KMST_gen-2002, KLNS-2002, kozlowski08} using a multiple-integral representation. The EFP has been considered also for the $6$-vertex model \cite{colomo08-1, colomo08-2, colomo08-3}, for higher spins XXZ \cite{highspinXXZ} and for dimer models \cite{stolze09}. We also remark that high temperature expansions of the EFP for Heisenberg chains have been studied in \cite{tsuboi-2005,tsuboi-2007,gohmann-2005}. A recent review of the EFP can be found in \cite{Fabio-Thesis} or \cite{abanov05}.

Field theory approaches are normally most suited for the calculation of large distance asymptotics of correlation functions, but conventional techniques like those inspired by the Luttinger Liquid paradigm (i.e. bosonization) are not appropriate for extended objects like the EFP and only capture its qualitative behavior, while being quantitatively unreliable, as it was showed in \cite{abanovkor}. The reason for this failure is that Luttinger Liquid is applicable only to low-energy excitations around the Fermi points, where the linear spectrum approximation is valid, while correlators like the EFP involve degrees of freedom very deep in the Fermi sea, where the whole spectrum with its curvature is important.

For this reason, the field theory calculation of the EFP requires a non-linear generalization of conventional bosonization, i.e. a true hydrodynamic description of the system. In \cite{abanov05} it was shown that, with such a non-linear collective description available, the calculation of the EFP is possible by employing, for instance, an instanton approach.

In this paper, we will extend the machinery developed in \cite{abanov05} and apply it to the spin-Calogero Model (sCM), for which a (gradientless) hydrodynamic description was recently constructed from its Bethe Ansatz solution \cite{KFA09}. The sCM is the spin$-1/2$ generalization \cite{1992-Polychronakos-exchange, 1992-HaHaldane, 1993-HikamiWadati} of the well-known Calogero-Sutherland model \cite{Sutherland-book} and is defined by the Hamiltonian
\begin{equation}
   H=-\frac{\hbar^{2}}{2}\sum_{j=1}^{N}\frac{\partial^{2}}{\partial x_{j}^{2}}
   +\frac{\hbar^{2}}{2}\left(\frac{\pi}{L}\right)^{2}\sum_{j\neq l}
   \frac{\lambda(\lambda-{\bf P}_{jl})}{\sin^{2}\frac{\pi}{L}\left(x_{j}-x_{l}\right)} \; ,
 \label{eq:h_afm}
\end{equation}
where ${\bf P}_{jl}$ is the operator that exchanges the positions of particles $j$ and $l$. We chose to analyze this Hamiltonian assuming it acts on fermionic particles, which means that the exchange term selects an anti-ferromagnetic ground state \cite{KFA09}. The coupling parameter $\lambda$ is taken to be positive and $N$ is the total number of particles.

In \cite{KFA09}, a collective description of the model was derived using four hydrodynamic fields: the density of particle with spin up/down $\rho_{\uparrow,\downarrow}$ and their velocities $v_{\uparrow, \downarrow}$. The Hamiltonian in terms of these fields is valid only for slowly evolving configurations, where terms with derivatives of the density fields can be neglected. This description is referred to as a {\it gradientless hydrodynamics}. In \cite{KFA09}, this theory was used to show the non-linear coupling between the spin and charge degrees of freedom beyond the Luttinger Liquid paradigm and it was shown that, while a charge excitation can evolve without affecting the spin sector (for instance for a spin singlet configuration), a spin excitation carries also some charge with it, in a non-trivial way.

The EFP for the sCM has not been considered in the literature yet. For the spin-less case of the Calogero-Sutherland interaction, the asymptotic behavior of the EFP was obtained using the form of the ground state wavefunction and thermodynamical arguments \cite{mehta91} (see \cite{Fabio-Thesis, abanov05} for details). It should be noted that for certain special values of the coupling parameter $\lambda$, the spin-less theory is tightly linked with Random Matrix Theory (RMT) and the EFP is the probability of having no energy eigenvalues in a given interval. For these values of $\lambda$ the EFP can be calculated with much greater accuracy due to the additional structure provided by RMT \cite{cloizeaux73}.

If we write the ground state of the system as $\Psi_G (x_1, x_2, \ldots , x_N)$, the Emptiness Formation Probability is defined as
\be
  P(R) \equiv {1 \over \langle \Psi_G | \Psi_G \rangle}
  \int_{|x_j|>R} \de x_1 \ldots \de x_N
  \left| \Psi_G (x_1, \ldots , x_N) \right|^2 \; ,
\ee
or, following \cite{QISM}
\be
  P(R) = \lim_{\alpha \to \infty} \langle \Psi_G |
  \eu^{-\alpha \int_{-R}^R \rho_c (x) \de x} | \Psi_G \rangle \; ,
\ee
where $\rho_c (x)$ is the total particle density operator
\be
   \rho_c (x) \equiv \sum_{j=1}^N \delta (x-x_j) \; .
\ee

For a model like the sCM, we can also introduce the EFPs for particles with spin up or down separately
\be
  P_{\uparrow,\downarrow} (R) = \lim_{\alpha \to \infty} \langle \Psi_G |
  \eu^{-\alpha \int_{-R}^R \rho_{\uparrow,\downarrow} (x) \de x} | \Psi_G \rangle \; ,
  \label{EFPs}
\ee
which will allow us to discuss the EFP as well as the PFFS.

The approach we use to calculate the EFPs (\ref{EFPs}) in this work is similar to what was explained in \cite{abanov05}. The idea is to consider the system as a quantum fluid evolving in imaginary time (Euclidean space). Then the EFP can be considered as the probability of a rare fluctuation that will deplete the region $-R<x<R$ of particle at a given imaginary time (say $\tau=0$). With exponential accuracy, the leading contribution to this probability comes from the action calculated on the saddle point solution (instanton) satisfying the EFP boundary condition.

In section \ref{sec:Two-fluid-picture} we will first review the results of \cite{KFA09} and transform them into an intriguing form where the dynamics can be decoupled into two independent fluids of spin-less Calogero-Sutherland particles. This two-fluid description is one of the interesting observations of this paper. In section \ref{sec:instantact} we will explain the instanton approach and formulate the problem in this language. In section \ref{sec:DFP} we will concentrate on a generalization of the EFP, the Depletion Formation Probability (DFP) which was introduced in \cite{abanovkor}. This correlator will allow us to calculate the different EFPs very efficiently by taking its different limits in section \ref{sec:EFP}. Most noticeably, we will derive the PFFS for the Haldane-Shastry model as the freezing limit of the sCM. In section \ref{sec:SDP} and \ref{sec:CDP} we will consider two additional DFP problems.
Instead of specifying boundary conditions for both the spin and charge sectors of the fluid as we did in the previous sections, we will now relax these conditions and constrain only one component at a time: this analysis suggests that an effective spin-charge separation can be conjectured for the EFP/DFP of the sCM. In section \ref{sec:discussion} we combine all these results and suggest a physical interpretation of them. The final section contains some concluding remarks. To avoid interruptions in the exposition, certain technical formalities are moved to the appendices and are organized as follows. In appendix \ref{ActionApp} we will revise and adapt the calculation of \cite{abanov05} to calculate the instanton action for our cases. In appendix \ref{sec:bosonization} we will repeat this calculation in the linearized hydrodynamics approximation or bosonization, to aid the discussions in section \ref{sec:discussion}.

\section{Two-fluid description}
\label{sec:Two-fluid-picture}

In \cite{KFA09} the gradientless hydrodynamic description for the sCM (\ref{eq:h_afm}) was derived in terms of densities and velocities of spin up and down particles: $\rho_{\uparrow,\downarrow} (t,x)$, $v_{\uparrow,\downarrow} (t,x)$. Here, we prefer to use densities and velocities of the {\it majority} and {\it minority} spin: $\rho_{1,2} (t,x), v_{1,2} (t,x)$, i.e. the subscript $1$ ($2$) takes the value $\uparrow$ or $\downarrow$ which ever is most (least) abundant species:
\begin{eqnarray}
  \rho_{1} & \equiv & \frac{\rho_{\uparrow}+\rho_{\downarrow}+|\rho_{\uparrow}-\rho_{\downarrow}|}{2}
  = \frac{\rho_c + \rho_s}{2} \; ,
  \label{rho1def}\\
  \rho_{2} & \equiv & \frac{\rho_{\uparrow}+\rho_{\downarrow}-|\rho_{\uparrow}-\rho_{\downarrow}|}{2}
  = \frac{\rho_c - \rho_s}{2} \; ,
  \label{rho2def}
\end{eqnarray}
where we introduced the charge and spin density
\bea
   \rho_c (t,x) & = & \rho_{\uparrow}+\rho_{\downarrow}
   = \rho_{1}+\rho_{2} \; ,
   \label{eq:rhocdef} \\
   \rho_s (t,x) & = & |\rho_{\uparrow}-\rho_{\downarrow}|
   =\rho_{1}-\rho_{2} \; .
   \label{eq:magnetization}
\eea
Please note that whatever species is majority or minority is decided dynamically in each point in space and time.
%and that the spin density is just the absolute value of the magnetization $m(x,t)$:
%\be
%   \rho_s (t,x) = |m(x,t)| \; .
%\ee

Under the condition \cite{KFA09}
\be
   \left| v_1 - v_2 \right| < \pi \rho_s \; ,
   \label{COCondition}
\ee
the Hamiltonian is
\begin{equation}
   H=\frac{1}{12\pi\left(\lambda+1\right)}\int_{-\infty}^{+\infty} \de x
   \left\{ k_{R1}^{3}-k_{L1}^{3}
   + \frac{1}{2\lambda+1} \left(k_{R2}^{3}-k_{L2}^{3}\right) \right\} \; ,
   \label{eq:hamiltonianCO}
\end{equation}
where
\begin{eqnarray}
   k_{R1,L1} & \equiv & v_{1} \pm (\lambda + 1) \pi  \rho_{1}
   \pm \lambda \pi \rho_2 \; ,
   \nonumber \\
   k_{R2,L2} & = & (\lambda + 1) v_{2} - \lambda v_1
   \pm (2 \lambda +1) \pi\rho_{2}
   \label{eq:kdef}
\end{eqnarray}
are the four {\it dressed Fermi momenta}.

It turns out that an auxiliary set of hydrodynamic variables decouples the Hamiltonian (\ref{eq:hamiltonianCO}) into the sum of two independent spin-less Calogero-Sutherland fluids $a$ and $b$:
\be
   H = H_a + H_b
   = \sum_{\alpha=a,b} \int \de x
   \left[ \frac{1}{2} \rho_\alpha v_\alpha^2
   + \frac{\pi^{2}\lambda_\alpha^2}{6} \rho_\alpha^3 \right] \; ,
   \label{eq:independentfluid}
\ee
where
\bea
   \rho_a & \equiv &
   \frac{k_{R1}-k_{L1}}{2\pi\lambda_a}
   = \rho_{1}+\frac{\lambda}{\lambda+1}\rho_{2} \; ,
   \label{tilderho1} \\
   \rho_b & \equiv & \frac{k_{R2}-k_{L2}}{2\pi\lambda_b}
   = \frac{1}{\lambda+1}\rho_{2} \; ,
   \label{tilderho2} \\
   v_a & \equiv &
   \frac{k_{R1}+k_{L1}}{2} = v_{1} \; ,
   \label{tildev1} \\
   v_b & \equiv &
   \frac{k_{R2}+k_{L2}}{2}
   = \left(\lambda+1\right)v_{2}-\lambda v_{1} \; ,
   \label{tildev2}\\
   \lambda_a & \equiv & \lambda+1\;,
   \label{beta}\\
   \lambda_b & \equiv & \left(\lambda+1\right)\left(2\lambda+1\right) \; .
   \label{gamma}
\eea

We remark that both the auxiliary variables and the real variables satisfy the canonical commutation relations, i.e.
\begin{equation}
   \left[ \rho_\alpha (x) , v_\beta (y) \right]
   = - \ii \hbar \delta_{\alpha,\beta} \delta^{\prime}(x-y) \; ,
   \qquad \alpha, \beta= \{1,2\}; \{a,b\} \; .
   \label{commrel}
\end{equation}
The form of the Hamiltonian (\ref{eq:independentfluid}) is one of the interesting observation of this paper, since it allows us to reduce the spin Calogero-Sutherland model into a sum of two spin-less theories. Each of the terms in square brackets in (\ref{eq:independentfluid}) is the gradientless Hamiltonian of a spin-less CS system with coupling constants $\lambda_{a,b}$ given by (\ref{beta}, \ref{gamma}). In \cite{abanov05} the gradientless hydrodynamics of spin-less particles, like the ones in (\ref{eq:independentfluid}) was used to calculate the EFP from the asymptotics of an instanton solution. In the next section we review this approach and we leave the mathematical details to appendix \ref{ActionApp}.

\section{The instantonic action}
\label{sec:instantact}

Let us perform a Wick's rotation to work in imaginary time $\tau \equiv \ii t$. Note that this makes the velocities in (\ref{eq:kdef}) imaginary ($v \rightarrow \ii v$) and the $k$'s complex numbers. The $x-t$ plane is mapped into the complex plane spanned by $z \equiv x + \ii \tau$.

Following \cite{abanov05}, we will calculate the EFP as an instanton configuration (i.e. a classical solution in Euclidean space) that satisfies the boundary condition
\be
   \rho_\alpha (\tau=0; -R<x<R) = 0 \; , \qquad \qquad \alpha=1,2,c \; ,
   \label{EFPbc}
\ee
in the limit $R \to \infty$, i.e. $R$ much bigger than any other length scale in the system. This limit guarantees that the gradient-less hydrodynamics (\ref{eq:independentfluid}) is valid in the bulk of the space-time. Once we have the classical solution of the equation of motion $\phi_{\rm EFP}$ that satisfies (\ref{EFPbc}), a saddle-point calculation gives the EFP with exponential accuracy as the action ${\cal S}$ calculated on this optimal configuration \cite{abanov05}:
\be
   P(R) \simeq \eu^{-{\cal S} \left[ \phi_{\rm EFP} \right] } \; .
\ee
Of course, to uniquely specify the problem, the boundary conditions at infinity have to be provided as well and we will take them to be those of an equilibrium configuration:
\begin{eqnarray}
    \rho_{1} (\tau,x) \stackrel{x,\tau \to \infty}{\rightarrow} \rho_{01} \; , & \quad\quad &
    v_{1} (\tau,x) \stackrel{x,\tau \to \infty}{\rightarrow} 0 \; ,
    \nonumber \\
    \rho_{2} (\tau,x) \stackrel{x,\tau \to \infty}{\rightarrow} \rho_{02} \; , & \quad\quad &
    v_{2} (\tau,x) \stackrel{x,\tau \to \infty}{\rightarrow} 0 \; .
    \label{eq:infinity_real_variables}
\end{eqnarray}
When $\rho_{01}=\rho_{02}$ we have an asymptotic singlet state (the AFM in zero magnetic field). The condition $\rho_{01}\neq\rho_{02}$ can be achieved via a constant external magnetic field which would result in a finite equilibrium magnetization. It is easy to implement these boundary conditions in our two-fluid description using (\ref{tilderho1}-\ref{tildev2}).

The key point for the calculation is that we can represent the hydrodynamic fields in terms of the dressed Fermi momenta $k_{R1},k_{R2}$ (which in Euclidean space become complex and complex conjugated to $k_{L1},k_{L2}$ respectively) through (\ref{eq:kdef}):
\be
   k_{R1} = \lambda_a \pi \rho_a + \ii v_a \; ,
   \qquad \qquad
   k_{R2} = \lambda_b \pi \rho_b + \ii v_b \; .
   \label{krhov}
\ee
In \cite{KFA09}, it was shown that these $k$-fields propagate independently according to 4 decoupled Riemann-Hopf equations
\be
  \partial_\tau w - \ii w \partial_x w = 0 \; ,
  \qquad \qquad w=k_{R,L;1,2} \; .
  \label{riemannhopf}
\ee
These equations have the general (implicit) solution
\be
   w = F(x + \ii w \tau)
   \label{implsol}
\ee
where $F(z)$ is an analytic function to be chosen to satisfy the boundary conditions.

Guided by \cite{abanov05}, the solution for an EFP problem is
\be
  k_{R1} = F_a (x + \ii k_{R1} \tau ) \; , \qquad \qquad
  k_{R2} = F_b (x + \ii k_{R2} \tau ) \; ,
  \label{ksols}
\ee
with
\begin{eqnarray}
   F_a (z) & \equiv & \lambda_a \pi \rho_{0a}
   + \lambda_a \pi \eta_a \left( {z \over \sqrt{z^2 - R^2}} - 1 \right) \; ,
   \label{eq:kr1}\\
   F_b (z) & \equiv & \lambda_b \pi \rho_{0b}
   + \lambda_b \pi \eta_b \left( {z \over \sqrt{z^2 - R^2}} - 1 \right) \; ,
   \label{eq:kr2}
\end{eqnarray}
which automatically satisfy the conditions at infinity (\ref{eq:infinity_real_variables}):
\begin{eqnarray}
   \rho_a (\tau,x \rightarrow \infty) & \rightarrow &
   \rho_{01} + \frac{\lambda}{\lambda+1}\rho_{02}
   \equiv \rho_{0a} \; ,
   \nonumber \\
   \rho_b (\tau,x \rightarrow \infty) & \rightarrow &
   \frac{1}{\lambda+1}\rho_{02} \; ,
   \equiv \rho_{0b} \; ,
   \nonumber \\
   v_{a,b} (\tau,x \rightarrow \infty) & \rightarrow & 0 \; ,
   \label{eq:auxiliaryBC0}
\end{eqnarray}
while $\eta_{a,b}$ are two, possibly complex, constants that allow to satisfy the EFP boundary conditions (\ref{EFPbc}).

In appendix \ref{ActionApp} we show that the instanton action can be expressed as a contour integral where only the behaviors of the solutions (\ref{ksols}) at infinity and close to the depletion region are needed, saving us the complication of solving the implicit equations in generality. Using the two-fluid description, the action can be written as the sum of two spin-less Calogero-Sutherland fluids: from (\ref{DFPactionResult}) we have
\be
   {\cal S}_{\rm EFP} = {1 \over 2} \, \pi^2 R^2
   \sum_{\alpha =a,b} \lambda_\alpha \, \eta_\alpha \, \bar{\eta}_\alpha \; .
   \label{EFPactionRes}
\ee

Before we proceed further, we should mention that the two-fluid description we employ is valid as long as the inequality (\ref{COCondition}) is satisfied. In fact, the solution (\ref{ksols}) could violate the inequality in a small region around the points $(\tau,x)=(0,\pm R)$. However, close to these points the hydrodynamic description is expected to be somewhat pathological, because gradient corrections (which we neglect) become important. As it was argued in \cite{abanovkor, abanov05}, the contributions that would come to the EFP from these small regions are subleading and negligible, in the asymptotic limit $R \to \infty$ we consider. Therefore, we do not need to worry about what happens near the points $(\tau,x)=(0,\pm R)$. However, a consequence of the ``singular'' nature of these points is that, in our solution, the species that constitutes the majority (minority) spin in the region of depletion $-R<x<R$ at $\tau=0$, could switch and become minority (majority) at infinity. This could be important to keep in mind in interpreting our formulae, but our formalism already takes that into account naturally.

\section{Depletion Formation Probability}
\label{sec:DFP}

It is more convenient to consider a generalization of the EFP problem, called Depletion Formation Probability (DFP) which was introduced in \cite{abanovkor}. In hydrodynamic language the DFP boundary conditions for the {\it majority} and {\it minority} spins are
\begin{eqnarray}
   \rho_{1}(\tau = 0;-R<x<R) & = &\tilde{\rho}_{1} \; ,
   \nonumber \\
   \rho_{2}(\tau = 0;-R<x<R) & = & \tilde{\rho}_{2} \; .
   \label{eq:dfp_real_variables}
\end{eqnarray}
The DFP is a natural generalization of the EFP (\ref{EFPbc}) and it reduces to it for $\tilde{\rho}_{1,2} =0$. Of course, there is some ambiguity on the microscopic definition of the DFP (see \cite{abanovkor, abanov05}). One can, for instance, consider it as the macroscopic version of the $s$-EFP introduced in \cite{keating06}. We will first calculate the DFP as the most general case and later take the appropriate interesting limits.

In terms of the auxiliary fields we introduced in (\ref{tilderho1}-\ref{tildev2}) to achieve the two-fluids description (\ref{eq:independentfluid}), the DFP boundary conditions are
\begin{eqnarray}
   \rho_a (\tau = 0,-R<x<R) & = &
   \tilde{\rho}_{1}+\frac{\lambda}{\lambda+1}\tilde{\rho}_{2}
   \equiv \tilde{\rho}_a \; ,
   \nonumber \\
   \rho_b (\tau = 0,-R<x<R) & = &
   \frac{1}{\lambda+1}\tilde{\rho}_{2}
   \equiv \tilde{\rho}_b \; .
   \label{eq:auxiliaryBC1}
\end{eqnarray}

We specify the parameters $\eta_{1,2}$ in (\ref{eq:kr1},\ref{eq:kr2}) by expressing the boundary conditions (\ref{eq:auxiliaryBC1}) in terms of the dressed momenta using
\begin{eqnarray}
   \rho_{1} & = & \frac{1}{\pi(\lambda+1)} \left[ \R \, k_{R1}
   - \frac{\lambda}{2\lambda+1} \, \R \, k_{R2} \right] \; ,
   \nonumber \\
   \rho_{2} & = & \frac{1} {\pi(2\lambda+1)} \, \R \, k_{R2} \; ,
   \nonumber \\
   v_{1} & = & \I \, k_{R1} \; ,
   \nonumber \\
   v_{2} & = & \frac{1}{\lambda+1} \Big[ \I \, k_{R2}
   + \lambda \, \I \, k_{R1} \Big] \; .
   \label{eq:DFPinstanton}
\end{eqnarray}
This leads to
\begin{eqnarray}
   \lambda_a \eta_a & = &
   \lambda_a \left( \rho_{0a} - \tilde{\rho}_a \right)
   \nonumber \\
   & = & (\lambda+1) (\rho_{01} - \tilde{\rho}_1 )
   + \lambda ( \rho_{02} - \tilde{\rho}_{2}) \; ,
   \nonumber \\
   \lambda_b \eta_b & = &
   \lambda_b \left( \rho_{0b} - \tilde{\rho}_b \right)
   \nonumber \\
   & = & ( 2 \lambda + 1 )  \left(\rho_{02} - \tilde{\rho}_{2} \right) \; .
   \label{generalDFPsol}
\end{eqnarray}

It is now straightforward to obtain the DFP by substituting (\ref{generalDFPsol}) into (\ref{EFPactionRes}). After some simple algebra we get
\begin{eqnarray}
   P_{DFP}(R)
%   & = & \exp \Biggl\{
%   - \frac{\left(\lambda+1\right)}{2} \pi^2 \left[ \rho_{01} -\tilde{\rho}_{1}
%   + \frac{\lambda}{\lambda+1} (\rho_{02}-\tilde{\rho}_{2}) \right]^{2} R^2
%   \nonumber \\
%   && \qquad - \frac{\left(2\lambda+1\right)} {2\left(\lambda+1\right)}
%   \pi^2 \left(\rho_{02}-\tilde{\rho}_{2}\right)^2 R^2 \Biggr\}
%   \nonumber \\
   & = & \exp \Biggl\{- {\pi^2 \over 2} \left[
   \lambda \left( \rho_{0c} - \tilde{\rho}_c \right)^2
   + \left( \rho_{01} - \tilde{\rho}_1 \right)^2
   + \left( \rho_{02} - \tilde{\rho}_2 \right)^2 \right] R^2 \Biggr\}
   \nonumber \\
   & = &\exp \Biggl\{- {\pi^2 \over 2} \left[
   (\lambda+\frac{1}{2}) \left( \rho_{0c} - \tilde{\rho}_c \right)^2
   + \frac{1}{2}\left( \rho_{0s} - \tilde{\rho}_s \right)^2
   \right] R^2 \Biggr\} \; ,
   \label{eq:generalDFP}
\end{eqnarray}
where we introduced a notation in terms of the charge field ($\rho_{0c}=\rho_{01}+\rho_{02}$, $\tilde{\rho}_c = \tilde{\rho}_1 + \tilde{\rho}_2$) and of the spin field ($\rho_{0s}=\rho_{01}-\rho_{02}$, $\tilde{\rho}_s = \tilde{\rho}_1 - \tilde{\rho}_2$).
Equation (\ref{eq:generalDFP}) is the main result of this work. To understand it better, we will consider several interesting limits.

\subsection{Asymptotic singlet state}

If no external magnetic field is applied, the equilibrium configuration of an anti-ferromagnetic system like the one we consider is in a singlet state. This means that in the boundary conditions at infinity (\ref{eq:infinity_real_variables}) we should set $\rho_{01}=\rho_{02}=\rho_{0}$. In this limit (\ref{eq:generalDFP}) reduces to
\begin{eqnarray}
   P_{DFP}^{\mbox{singlet}} (R) & = &
   \exp \Biggl\{- {\pi^2 \over 2} \left[
   \lambda \left( 2\rho_{0} - \tilde{\rho}_c \right)^2
   + \left( \rho_{0} - \tilde{\rho}_1 \right)^2
   + \left( \rho_{0} - \tilde{\rho}_2 \right)^2 \right] R^2 \Biggr\}
   \nonumber \\
   & = & \exp \Biggl\{- {\pi^2 \over 2} \left[
   (\lambda+\frac{1}{2}) \left( 2\rho_{0} - \tilde{\rho}_c \right)^2
   + \frac{1}{2}\tilde{\rho}_s^2 \right] R^2 \Biggr\} \; .
   \label{eq:singletDFP}
\end{eqnarray}

\section{Emptiness Formation Probability}
\label{sec:EFP}

By taking the limit $\tilde{\rho}_{1,2}=0$ we can use (\ref{eq:generalDFP}) to calculate the different EFPs. The probability to find the region $-R < x < R$ at $\tau=0$ completely empty of particles is therefore
\bea
   P_{EFP}(R) & = & \exp \Biggl\{- {\pi^2 \over 2} \left[
   \lambda \left( \rho_{01} + \rho_{02} \right)^2
   + \rho_{01}^2 + \rho_{02}^2 \right] R^2 \Biggr\}
   \nonumber \\
   & = & \exp \Biggl\{- {\pi^2 \over 2} \left[
   (\lambda+\frac{1}{2}) \rho_{0c}^2
   + \frac{1}{2} \rho_{0s}^2 \right] R^2 \Biggr\} \; ,
   \label{eq:EFP}
\eea
which becomes
\be
   P_{EFP}^{\mbox{singlet}} (R) =
   \exp \biggl\{- \frac{\pi^2}{2} (\lambda + \frac{1}{2}) (2\rho_0 )^2 R^2 \biggr\}
   \label{eq:EFP-singlet}
\ee
for the asymptotic singlet state. This is equivalent to the EFP of a spin-less Calogero-Sutherland system with coupling constant $\lambda' = \lambda + 1/2$, see (\ref{eq:CS-EFP}). This is consistent with the phase-space picture provided in \cite{KFA09}, in which it is explained that for a singlet state each particle occupies an area of $\pi(\lambda +1/2)$ due to the exclusion statistics, while it would occupy an area $\pi(\lambda +1)$ if it were alone. Therefore, in this context, the charge field can be thought of as describing a spin-less Calogero system with coupling constant $\lambda' = \lambda + 1/2$.

\subsection{Free fermions with spin}

Setting the coupling parameter $\lambda=0$ corresponds to non-interacting (free) fermions with spins and this reduces (\ref{eq:generalDFP}) to
\bea
   P_{DFP}^{\mbox{free fermions}}(R)
   & = & \exp\Biggl\{
   -\frac{1}{4}\left[\pi\left(\rho_{0c}-\tilde{\rho}_{c}\right)R\right]^{2}
   -\frac{1}{4}\left[\pi\left(\rho_{0s}-\tilde{\rho}_{s}\right)R\right]^{2}\Biggr\}
   \nonumber \\
   & = & \exp\Biggl\{
   -\frac{1}{2}\left[\pi\left(\rho_{01}-\tilde{\rho}_{1}\right)R\right]^{2}
   -\frac{1}{2}\left[\pi\left(\rho_{02}-\tilde{\rho}_{2}\right)R\right]^{2}\Biggr\}
   \; .
   \nonumber \\
   \label{eq:FFDFP}
\eea
This result is the same as the one obtained in \cite{abanov05}.
The EFP is then
\begin{equation}
   P_{EFP}^{\mbox{free fermions}}(R) = \exp\Biggl\{
   -\frac{1}{2}\left(\pi \rho_{01} R\right)^{2}
   -\frac{1}{2}\left(\pi \rho_{02} R\right)^{2}\Biggr\} \; ,
   \label{eq:FFEFP}
\end{equation}
which agrees with the results obtained in the context of Random Matrix Theory \cite{cloizeaux73,mehta91}, where the subleading corrections were also found.

\subsection{Spin-less Calogero-Sutherland model}

Of course, the prime check to our formula for the DFP/EFP of the sCM is to take its spin-less limit $\tilde{\rho}_2 = \rho_{02} = 0$, which gives
\bea
   P_{DFP}^{\mbox{spin-less}}(R) & = & \exp \Biggl\{
   - \frac{\left(\lambda+1\right)}{2} \pi^2 \left( \rho_{01} -\tilde{\rho}_{1}
   \right)^{2} R^2 \Biggr\} \; ,
   \label{eq:CS-DFP} \\
   P_{EFP}^{\mbox{spin-less}}(R) & = & \exp \Biggl\{
   - \frac{\left(\lambda+1\right)}{2} \pi^2 \rho_{01}^2 R^2 \Biggr\} \; ,
   \label{eq:CS-EFP}
\eea
in perfect agreement with \cite{abanov05, mehta91} for a spin-less Calogero-Sutherland system with coupling $\lambda' = \lambda +1$.
%In terms of a single particle picture one can say that the (\ref{eq:CS-EFP}) indicates that each particle occupies a phase space volume ($\lambda+1$) if there are no particles of the other species present in this volume. Interesting discussion on fractional exclusion statistics in sCM can be found in \cite{KFA09} where it was shown that the single-particle phase space picture for sCM admits a much more interesting form as compared to its spin-less counterpart.

\subsection{Probability of Formation of Ferromagnetic Strings}

If we require the minority spin particles to completely empty the region $-R<x<R$ at $\tau=0$, we are left only with the majority spin and we created a (partially) polarized state. We can refer to this case as the Probability of Formation of Partially Ferromagnetic Strings (PFPFS) \cite{abanovkor, abanov05}. Setting $\tilde{\rho}_{2}=0$ in (\ref{eq:generalDFP}) and leaving $\tilde{\rho}_{1}$ finite we have
\begin{equation}
   P_{PFPFS} (R) = \exp \Biggl\{- {\pi^2 \over 2} \left[
   \lambda \left( \rho_{01} - \tilde{\rho}_1 + \rho_{02} \right)^2
   + \left( \rho_{01} - \tilde{\rho}_1 \right)^2
   + \rho_{02}^2 \right] R^2 \Biggr\} \; .
   \label{eq:ferrostrings}
\end{equation}
The above is the probability of formation of ferromagnetic strings accompanied by a partial depletion of particles, since in the region of depletion we have $\tilde{\rho}_{c}=\tilde{\rho}_{1}$. We can impose that the average density of particles is constant everywhere by setting $\tilde{\rho}_{1}=\rho_{01}+\rho_{02}=\rho_{0c}$, while still requiring all particles in the region $-R<x<R$ at $\tau=0$ to be completely polarized (maximal magnetization: PFFS)
\begin{equation}
   P_{PFFS} (R) = \exp \biggl\{
   -\left[\pi\rho_{02}R\right]^{2}\biggr\} \; .
   \label{eq:frozen_ferro_DFP}
\end{equation}
Note that (\ref{eq:frozen_ferro_DFP}) is independent of $\lambda$ and exactly corresponds to the Emptiness Formation Probability of a $\lambda' = \lambda +1 =2$ spin-less Calogero model with background density given by $\rho_{02}$ (\ref{eq:CS-EFP}). Interestingly the same result (\ref{eq:hsmdfp}) will be derived in the next sections as the EFP of minority spins, i.e. $\tilde{\rho}_{2}=0$, in the Haldane-Shastry model (\ref{eq:Haldane_shastry}). This is just another aspect of the well-known relation between spin-Calogero, Haldane-Shastry and $\lambda'=2$ spin-less Calogero models \cite{1988-Haldane-HS, KFA09} as it will be shown in the next section.

\subsection{The freezing limit}
\label{sec:HSM}

If we take the $\lambda\rightarrow\infty$ limit in the spin-Calogero model (\ref{eq:h_afm}), the charge dynamics freezes (the particles become pinned to a lattice) and only the spin dynamics survives. This {\it freezing limit} was shown by Polychronakos \cite{1993-Polychronakos} to be equivalent to the Haldane-Shastry model (HSM) \cite{1988-Haldane-HS, 1988-Shastry-HS}:
\begin{equation}
   H_{\rm HSM}= 2 {\pi^2 \over N^2} \sum_{j<l}
   \frac{S_{j} \cdot S_{l}}{\sin^2 {\pi \over N} \left(j-l\right)} \; ,
   \label{eq:Haldane_shastry}
\end{equation}
an integrable Heisenberg chain with long range interaction. In \cite{KFA09} the freezing limit was studied through a systematic expansion of the hydrodynamic fields in inverse powers of $\mu \equiv \lambda+1/2$:
\be
   \Omega = \Omega^{(0)} + \frac{1}{\mu} \Omega^{(1)}
   + \frac{1}{\mu^{2}} \Omega^{(2)} + \ldots \; ,
   \qquad \qquad
   \rho_{c}, v_{c}, \rho_{s}, v_{s} \rightarrow \Omega \; .
\ee
With an additional rescaling of time $t \to \mu t$, the equations of motion were separated order by order in powers of $\mu$.

It was shown that the charge sector is frozen, in that charge dynamics appears only at orders $\Ord (\mu^{-1})$ and higher, while the spin sector already has non-trivial dynamics at order $\Ord (1)$. This dynamics is the same as the one derived independently for the HSM, over a background density of particles $\rho_{0c} = N / L$. As a consequence of charge freezing, this background density is kept fixed and constant up to order $\Ord (1)$ and fluctuation are suppressed as $1 / \mu$.
Therefore, as $\mu = \lambda + 1/2 \to \infty$, charge conservation is imposed dynamically everywhere, including in the region of depletion:
\begin{equation}
   \tilde{\rho}_c^{(0)} = \tilde{\rho}_1^{(0)} + \tilde{\rho}_2^{(0)}
   = \rho_{01} + \rho_{02} = \rho_{0c} \; .
   \label{eq:finiteP}
\end{equation}

The above (\ref{eq:finiteP}) along with the usual depletion boundary conditions (\ref{eq:dfp_real_variables}) reduces (\ref{eq:generalDFP}) to
\bea
   P_{DFP}^{\mu\rightarrow\infty} (R) & = &
   \exp \biggl\{ - {\pi^2 \over 2}
   \left[{1 \over 2} \left(\rho_{0s}-\tilde{\rho}_{s}^{(0)}\right)^2
   + {1 \over \mu} \tilde{\rho}_c^{(1)} + \Ord (\mu^{-2}) \right]
   R^{2} \biggr\}
   \label{eq:muinfdfp} \\
   & \simeq & \exp \biggl\{
   -\left[\pi\left(\rho_{01}-\tilde{\rho}_{1}^{(0)}\right)R\right]^{2}\biggr\}
   = \exp \biggl\{
   -\left[\pi\left(\rho_{02}-\tilde{\rho}_{2}^{(0)}\right)R\right]^{2}\biggr\} \; , \nonumber
\eea
where corrections for a finite $\lambda$ are of the order $1 / \mu$. Eq. (\ref{eq:muinfdfp}) coincides with (\ref{eq:frozen_ferro_DFP}), where condition (\ref{eq:finiteP}) was imposed as a boundary condition, and not dynamically from the equations of motion.

\subsection{Haldane-Shastry model}
\label{HSM-independent}

The hydrodynamic description of the HSM (\ref{eq:Haldane_shastry}) was constructed in \cite{KFA09}, resulting in the following Hamiltonian for the minority spins (remember that the HSM is a lattice model and therefore there is no charge dynamics)
\bea
   H_{\rm HSM} & = &
    \int \de x \left[ {1 \over 2} \; \rho_2 \; v_2^2
   + {2 \over 3} \; \pi^2 \; \rho_2^3 \right] \; ,
   \label{H-HSM}
\eea
which is the hydrodynamic Hamiltonian for a spin-less Calogero system with coupling constant $\lambda' = 2$, see (\ref{eq:independentfluid}). It is in fact known that the spectrum of the HSM is equivalent to that of a spin-less Calogero-Sutherland model with exclusion parameter $\lambda'=2$, but with a high degeneracy due to the underlying Yangian symmetry \cite{1988-Haldane-HS}.

The connection between the HSM (\ref{H-HSM}) and the sCM in the freezing limit is to express the minority spin fields in (\ref{H-HSM}) in terms of spin fields \cite{KFA09}:
\be
   \rho_2 =\frac{\rho_{0c} -\rho_{s}}{2} \, , \qquad \qquad
   v_2 = - 2v_{s}\, .
 \label{frid}
\ee

It is straightforward, using (\ref{eq:CS-DFP}) with $\lambda'=2$, to see that the PFPFS for the HSM model is exactly (\ref{eq:muinfdfp}):
\be
   P_{PFPFS}^{HSM} (R) = \exp \biggl\{
   -\left[\pi\left(\rho_{02}-\tilde{\rho}_{2}\right)R\right]^{2}\biggr\} \; .
   \label{eq:hsmdfp}
\ee
The equivalence between (\ref{eq:frozen_ferro_DFP}), (\ref{eq:muinfdfp}) and (\ref{eq:hsmdfp}) is a strong check of the consistency of our methods and shows from a novel perspective the well-known relations between the sCM in the large $\lambda$ limit, the HSM and the spin-less Calogero-Sutherland model with $\lambda' =2$.

\section{Spin Depletion Probability}
\label{sec:SDP}

So far, we considered a DFP problem specified by the boundary conditions (\ref{eq:dfp_real_variables}), i.e. by fixing the density of both species of particles on the segment $\tau=0$, $-R<x<R$. However, our formalism allows for a more general and natural question. We can, for instance, demand a given magnetization (i.e. spin density) on the segment, without constraining the charge sector, i.e. imposing the boundary condition:
\be
   \rho_s (\tau=0; -R<x<R) = \tilde{\rho}_s \; ,
   \label{rhosbc}
\ee
instead of (\ref{eq:dfp_real_variables}).

From (\ref{eq:DFPinstanton}) we have
\be
  \rho_s = {1 \over \pi (\lambda +1)} \R \left[ k_{R1} - k_{R2} \right] \; ,
\ee
substituting in (\ref{ksols},\ref{eq:kr1},\ref{eq:kr2}) we find that (\ref{rhosbc}) is satisfied if
\be
   \lambda_a \eta_a - \lambda_b \eta_b
   = \lambda_a \rho_{0a} - \lambda_b \rho_{0b} - (\lambda+1) \tilde{\rho}_s \; .
\ee
This equation leaves undetermined a complex constant $\xi = \xi_1 + \ii \xi_2$: for later convenience we parametrize the solution as
\bea
   \lambda_a \eta_a & = &
   \lambda_a \rho_{0a} - {1 \over 2} \tilde{\rho}_s
   - \left( \lambda + {1 \over 2} \right) \xi
   \nonumber \\
   & = & \left( \lambda + {1 \over 2} \right) \left( \rho_{0c} - \xi \right)
   + {1 \over 2} \left( \rho_{0s} - \tilde{\rho}_s \right) \; ,
   \nonumber \\
   \lambda_b \eta_b & = & \lambda_b \rho_{0b}
   + \left( \lambda + {1 \over 2} \right) \left( \tilde{\rho}_s - \xi \right)
   \nonumber \\
   & = & \left( \lambda + {1 \over 2} \right)
   \left( \rho_{0c} - \xi - \rho_{0s} + \tilde{\rho}_s \right) \; .
   \label{SpinEFPetas}
\eea
We have constructed the solution that realizes a constant spin density in the depletion region, while leaving the densities for the individual species free to vary. Please note that a finite imaginary part of $\xi$ is necessary to have $\partial_x \rho_{1,2} (\tau=0; -R<x<R) \ne 0$.

We can now substitute (\ref{SpinEFPetas}) in (\ref{EFPactionRes}) to find:
\be
   P_{SDP} (R;\xi) = \exp \biggl\{ - {\pi^2 \over 2} \left[
   \left( \lambda + {1 \over 2} \right)
   \left[ \left( \rho_{0c} - \xi_1 \right)^2 + \xi_2^2 \right]
   + {1 \over 2} \left( \rho_{0s} - \tilde{\rho}_s \right)^2
   \right] R^2 \biggr\} \; .
   \label{SDPxi}
\ee
This probability depends on two, yet undetermined, parameters: $\xi_{1,2}$. If we choose $\xi_1 = \tilde{\rho}_c $ and $\xi_2=0$ we recover exactly (\ref{eq:generalDFP}), as we expected. This would correspond to forcing a given charge density at the depletion region, together with (\ref{rhosbc}).

We also note that the configuration that maximizes the probability (\ref{SDPxi}) is given by $\xi = \rho_{0c}$:
\be
   P^{\rm Max}_{SDP} (R)= P_{SDP} (R;\xi=\rho_{0c})
   = \exp \biggl\{ - {\pi^2 \over 4}
   \left( \rho_{0s} - \tilde{\rho}_s \right)^2 R^2 \biggr\} \; .
   \label{SDPMax}
\ee
One cannot help but noticing the similarity between (\ref{SDPMax}) and (\ref{eq:muinfdfp}) or (\ref{eq:hsmdfp}). 

Since the parametrization we choose in (\ref{SpinEFPetas}) allowed us to express the probability (\ref{SDPxi}) as a Gaussian for $\xi_{1,2}$, we would get the same result by performing an integral over the free parameters:
\be
   P_{SDP}^{\rm opt} (R) = \int_{-\infty}^\infty \de \xi_1 \int_{-\infty}^\infty \de \xi_2 \, P_{SDP} (R;\xi)
   = P^{\rm Max}_{SDP} (R) .
   \label{SDPopt}
\ee
where the prefactor coming from the Gaussian integration is beyond our accuracy anyway (note, however, that it does not depend on the charge density). The integration over $\xi$ corresponds to summing over all configurations of the form (\ref{ksols}, \ref{eq:kr1}, \ref{eq:kr2}). 

The probability of realizing the magnetization set by (\ref{rhosbc}) is given by a sum over all configurations that satisfy the given boundary conditions. To perform this sum correctly, we would need to consider all possible charge density profiles $\tilde{\rho}_c (x)$ at the depletion region and therefore consider more general solutions than (\ref{eq:kr1}, \ref{eq:kr2}). These general solutions are of the form
\begin{eqnarray}
   F_a (z) & \equiv & \pi \left[ \lambda_a \rho_{0a} - {1 \over 2} \tilde{\rho}_s \right] {z \over \sqrt{z^2 - R^2}} 
   + {\pi \over 2} \tilde{\rho}_s 
   + \pi \left( \lambda + {1 \over 2} \right) \xi (z) \; ,
   \\
   F_b (z) & \equiv & \pi \left[ \lambda_b \rho_{0b} + \left( \lambda + 
   {1 \over 2} \right) \tilde{\rho}_s \right] {z \over \sqrt{z^2 - R^2}} 
   + \pi \left( \lambda + {1 \over 2} \right) 
   \left[ -\tilde{\rho}_s + \xi (z) \right] \; ,
   \nonumber
\end{eqnarray}
where $\xi (z)$ is an analytic function such that $\R \xi(x) = \tilde{\rho}_c (x)$ and $\xi (z \to \infty) \to 0$. The sum over all configurations satisfying (\ref{rhosbc}) can be formulated as a functional integral over all functions $\xi(z)$. However, it is easy to convince oneself that the configurations that minimize the action are of the form (\ref{eq:kr1}, \ref{eq:kr2}), with $\xi=\rho_{0c}$.

\section{Charge Depletion Probability}
\label{sec:CDP}

The last problem we will address is conjugated to the one considered in the previous section, i.e. the probability of realizing a given depletion of the charge
\be
   \rho_c (\tau=0; -R<x<R) = \tilde{\rho}_c \; ,
   \label{rhocbc}
\ee
without constraining the spin density.
Using (\ref{eq:DFPinstanton}) we have
\be
  \rho_c = \R \left[ {k_{R1} \over \pi \lambda_a}
  + {k_{R2} \over \pi \lambda_b} \right] \; ,
\ee
which means that (\ref{ksols},\ref{eq:kr1},\ref{eq:kr2}) fulfill (\ref{rhocbc}) if
\be
   \eta_a + \eta_b
   = \rho_{0a} + \rho_{0b} - \tilde{\rho}_c \; .
\ee
Once again, we are left with the freedom of introducing a complex number $\xi = \xi_1 + \ii \xi_2$ to parametrize the solution:
\bea
   \eta_a & = & \rho_{0a} - {2\lambda + 1 \over 2(\lambda +1)} \tilde{\rho}_c
   - {1 \over 2(\lambda +1)} \xi
   \nonumber \\
   & = & {2 \lambda +1 \over 2(\lambda +1)}
   \left( \rho_{0c} -\tilde{\rho}_c \right)
   + {1 \over 2(\lambda +1)} \left( \rho_{0s} - \xi \right) \; ,
   \nonumber \\
   \eta_b & = & \rho_{0b} - {1 \over 2(\lambda +1)} \tilde{\rho}_c
   + {1 \over 2(\lambda +1)} \xi
   \nonumber \\
   & = & {1 \over 2(\lambda +1)} \left( \rho_{0c} - \tilde{\rho}_c - \rho_{0s} + \xi \right) \; .
\eea
Inserting this into (\ref{EFPactionRes}) we obtain:
\be
   P_{CDP} (R;\xi) = \exp \biggl\{ - {\pi^2 \over 2} \left[
   \left( \lambda + {1 \over 2} \right)
   \left( \rho_{0c} - \tilde{\rho}_c \right)^2
   + {1 \over 2} \left[ \left( \rho_{0s} - \xi_1 \right)^2 +
   \xi_2^2 \right] \right] R^2 \biggr\} \; .
   \label{CDPxi}
\ee
Setting $\xi_1 = \tilde{\rho}_s$ and $\xi_2 =0$ correctly reproduces (\ref{eq:generalDFP}), while the maximal probability is achieved for $\xi = \rho_{0s}$:
\be
   P^{\rm Max}_{CDP} (R) = P_{CDP} (R;\xi=\rho_{0s})
   = \exp \biggl\{ - {\pi^2 \over 2} \left[
   \left( \lambda + {1 \over 2} \right)
   \left( \rho_{0c} - \tilde{\rho}_c \right)^2 \right] R^2 \biggr\} \; .
   \label{CDPMax}
\ee
As before, since (\ref{CDPxi}) is Gaussian in $\xi_{1,2}$, we would obtain the same result by integrating over these variables
\be
   P_{CDP}^{\rm opt} (R) = \int_{-\infty}^\infty \de \xi_1 \int_{-\infty}^\infty \de \xi_2 \, P_{CDP} (R;\xi)
   = P^{\rm Max}_{CDP} (R) \; ,
   \label{CDPopt}
\ee
where we neglected the coefficient coming from the Gaussian integration because its beyond the accuracy of our methodology. This integration corresponds to summing over all configurations given by (\ref{eq:kr1}, \ref{eq:kr2}). Again, we notice a striking similarity between (\ref{CDPMax}, \ref{CDPopt}) and (\ref{eq:generalDFP}). We will comment in the next section on how to interpret these results.

\section{Discussion of the results}
\label{sec:discussion}

Eq. (\ref{eq:generalDFP}) looks like the product of the two independent depletion probabilities for the spin and charge sector. This interpretation is supported by the results of the two previous sections, see (\ref{SDPMax}) and (\ref{CDPMax}), but it is quite surprising in a sense. In fact, it would indicate a sort of an effective spin-charge separation, as if the spin and charge degrees of freedom could be depleted independently. This is contrary to intuition, since spin-charge separation is realized only for low-energy excitations close to the Fermi points, while the EFP involves degrees of freedom deep within the Fermi sea (and requires a full non-linear hydrodynamic description beyond the usual bosonization approach). However, it seems that from a EFP perspective spin-charge separation survives beyond the linearization of the spectrum, at least at leading order for the sCM.

In fact, quite surprisingly, for the Calogero-type interaction (as well as for free fermions), the DFP result obtained for small depletions using a linearized hydrodynamics (conventional bosonization) can be extended up to a complete emptiness and remain quantitatively correct \cite{abanov05}. This is due to the fact that the gradientless hydrodynamic for this interaction is purely cubic, see (\ref{eq:hamiltonianCO}). This fact has two important consequences: the first one is that the equations of motion can be written as Riemann-Hopf equations (\ref{riemannhopf}), which are trivially integrable with the implicit solution given by (\ref{implsol}). This is important to connect the boundary conditions at infinity with those due to the DFP.
The second fact is connected to the form of the parameters $u$ and $\kappa$ of the linearized theory, which in Hamiltonian formalism can be written in general as
\be
   H = \int \de x \left[ {u \over 2 \kappa} \Pi^2 + {u \, \kappa \over 2}
   \left(\nabla \phi \right)^2 \right] \; ,
\ee
where $\Pi(x)$ and $\phi(x)$ are conjugated fields. In terms of hydrodynamic variables (\ref{commrel}) they are
\be
   v (x) \equiv \Pi (x)\; , \qquad \qquad
   \rho (x)  \equiv \rho_0 + \nabla \phi (x) \; ,
\ee
where $\rho_0$ is the background value over which we are linearizing the theory. Note that $\kappa = {1 \over \pi K}$, where $K$ is the conventional Luttinger parameter \cite{giamarchi}.

For Calogero-type models, the sound velocity $u$ depends linearly on the density, while the interaction parameter $\kappa$ does not depend on the point around which we are linearizing. In appendix \ref{sec:bosonization} we show that the sound velocity can be rescaled out of the DFP calculation (at zero temperature) and $\kappa$ is the relevant factor encoding the interaction, which determines the coefficient of the Gaussian behavior of the DFP. All these peculiarities of the Calogero interaction conspire in a way that extending the small depletion result to higher depletion is ``trivial'' and, in fact, gives the correct result. Let us remark in this respect, that any non-linear theory can be seen as the integration of successive linear approximation, where the coefficients are adjusted at each point. In this light and from what we pointed out above, it is clear that the simplicity of the Calogero interaction allows a simple integration of successive linear theories for the DFP calculation and this is the reason for which the linearized result can be trivially extended from a small DFP to a complete EFP.

In appendix \ref{sec:bosonization} we calculate the DFP in the linearized approximation (guided by \cite{abanov05}). The result is (\ref{linearDFPres})
\be
   S_{DFP}^{\rm linear} = {\pi \over 2} \, \kappa \,
   (\rho_0 - \tilde{\rho})^2 \, R^2 \; ,
   \label{linearSDFP}
\ee
where we used $\eta = \bar{\eta} = \rho_0 - \tilde{\rho}$.

If we substitute (\ref{eq:kdef}) in (\ref{eq:hamiltonianCO}) we can write the hydrodynamic Hamiltonian in terms of spin and charge fields as
\bea
   H & = & \int \de x \left\{
   \frac{1}{2} \rho_{c} v_{c}^{2}
   + \frac{\pi^{2}}{6} \left( \lambda + {1 \over 2} \right)^2 \rho_{c}^{3}
   + \rho_{s} v_{c} v_{s} \right.
   \label{Hcs} \\
   && \qquad \left.
   + \left[ \left( \lambda + {1 \over 2} \right) \rho_{c} - \lambda \rho_{s}
   \right] v_{s}^{2}
   + \frac{\pi^{2}}{4} \left( \lambda + {1 \over 2} \right) \rho_{c} \rho_{s}^{2}	
   - \frac{\pi^{2}}{12} \, \lambda \, \rho_{s}^{3} \right\} \; .
   \nonumber
\eea
By linearizing this Hamiltonian, i.e. by expanding the fields as $\rho_{c,s} = \rho_{0;c,s} + \delta \rho_{c,s}$ and looking at the coefficients in front of the quadratic part, we find the following parameters:
\bea
    u_c (\rho_{0c}) = \pi \left( \lambda + {1 \over 2} \right) \rho_{0c} \; ,
    \quad && \quad
    \kappa_c (\rho_{0c}) = \pi \left( \lambda + {1 \over 2} \right) \; ,
    \\
    u_s (\rho_{0s}) = \pi \left( \lambda + {1 \over 2} \right) \rho_{0c}
    - \pi \lambda \rho_{0s} \; ,
    \quad && \quad
    \kappa_s (\rho_{0s}) = {\pi \over 2} \; .
\eea
We see that substituting these values in (\ref{linearSDFP}) correctly reproduce (\ref{SDPMax}, \ref{CDPMax}) and therefore (\ref{eq:generalDFP}) as well.
This means that not only the linearized theory is sufficient to calculate the correct coefficients of the EFP, but also that the spin-charge separation survives as if the linear theory was valid for high depletions as well.

To conclude, we can suggest a simple physical interpretation of (\ref{linearSDFP}). In a Calogero-Sutherland system, the interaction parameter $\kappa= \pi \lambda'$ has a simple semiclassical interpretation in terms of the phase-space area occupied by a single particle, see \cite{1997-KatoYamamotoArikawa} and \cite{KFA09}. We can then see that (\ref{linearSDFP}) represent a volume in the $x-\tau-k$ space: the phase-space area at a given $\tau$ is of the order of $\kappa (\rho_0 - \tilde{\rho}) R$, see, for instance, (\ref{krhov}). This has to be multiplied by the number of particles involved in the depletion over time, which is of the order $(\rho_0 - \tilde{\rho}) R$.

\section{Conclusions}
\label{sec:conclusions}

We calculated the Emptiness and Depletion Formation Probability for the spin Calogero-Model (\ref{eq:h_afm}) and for the Haldane-Shastry Model (\ref{eq:Haldane_shastry}). The EFP is one of the fundamental correlators in the theory of integrable models and, despite its being non-local, is considered to be one of the simplest. Nonetheless its asymptotic behavior is known only for a few systems and in this paper we calculated it for the sCM and the HSM for the first time, at the leading order.

The DFP is a natural generalization of the EFP in the hydrodynamics formalism we employ. By calculating the DFP in its most generality (\ref{eq:generalDFP}), we can achieve the different EFPs by taking its appropriate limit. The calculation is done in an instanton picture, where the DFP is viewed as the probability of formation of a rare fluctuation in imaginary time that realizes the required depletion at a given moment.

The long distance asymptotics in 1-D models are normally calculated in a field theory approach using bosonization. However, as this approach is valid only for low-energy excitations close to the Fermi points where the linearization of the spectrum is a reasonable approximation, it is not sufficient for the EFP, which involves degrees of freedom deep in the Fermi sea. For this reason we used a non-linear version of bosonization, i.e. the hydrodynamic description developed in \cite{KFA09}.

All our formulae show a characteristic Gaussian behavior as a function of the depletion radius $R$. This is to be expected for a gapless one-dimensional system, as it was first argued in \cite{abanovkor}. This is because in the asymptotic limit we consider, $R$ is the biggest length scale in the system and therefore the instanton configuration will have a characteristic area of $R^2$, where the second power comes from the dimensionality of the space-time. In \cite{abanovkor, abanov05} it was also shown that for small depletion, the linearized bosonization approach is sufficient to calculate the DFP, while in general it deviates from the correct results for progressively bigger depletion and, eventually, emptiness.

However, the Calogero-Sutherland kind of models (as well as non-interacting fermions) are special and the linearized result happen to coincide with the correct, non-linear one. We argued on the origin of this observation in the previous section. Moreover, we noticed that (\ref{eq:generalDFP}) and the analysis of section \ref{sec:SDP} and \ref{sec:CDP} indicates that, from a EFP perspective, spin-charge separation seems to survive beyond the linear approximation, in disagreement with what one na\"ively would expect. This resurgence of linear results in a non-linear problem is a very surprising result, peculiar of the sCM.

The coefficients in front of $R^2$ are novel of this work. In section \ref{sec:discussion} we interpreted them from a bosonization point of view and via a simple semiclassical argument and throughout the paper we have checked them against known results in certain limits where possible. In particular, we showed agreement with the free fermionic limit (\ref{eq:FFDFP}) and the spin-less Calogero-Sutherland model (\ref{eq:CS-DFP}). For both of these models, the EFP has a particular interest coming from Random Matrix Theory, as it is known that for certain rational values of the coupling parameter $\lambda$ the CSM describes the RMT ensembles. It would be interesting if the sCM would also have an interpretation in terms of some generalized random matrix model, but we are not aware of such connection yet.

In section \ref{sec:HSM} we used the fact that the Haldane-Shastry model can be achieved as the freezing limit ($\lambda \to \infty$) of the sCM to calculate the Probability of Formation of (Partially) Ferromagnetic Strings in the HSM. In section \ref{HSM-independent}, the same quantity was derived independently from the hydrodynamic description of the HSM. Section \ref{sec:HSM} and \ref{HSM-independent} highlight the correspondence between large-$\lambda$ sCM, HSM and $\lambda' = \lambda +1 =2$ spin-less Calogero model from a EFP/DFP perspective.
%In a significant part of the paper the DFP boundary conditions for both spin and charge were specified. However, in section \ref{sec:SDP} and \ref{sec:CDP}, the charge and spin boundary conditions were relaxed respectively. Apart from obtaining generalized results, this analysis suggested an effective spin-charge separation for the EFP/DFP in sCM and this was elaborated in section \ref{sec:discussion}.

\section*{Acknowledgments}

We are grateful to A.G. Abanov for suggesting us to consider the EFP problem for the sCM, for his notes on the spin-less problem and for several very useful discussions during this work. F.F. would like to thank A. Scardicchio for some interesting discussions. M.K. thanks H. Katsura who motivated us in employing our hydrodynamic approach towards the calculation of correlation functions. M.K. would like to acknowledge the hospitality of The Abdus Salam International Centre for Theoretical Physics, Trieste, Italy during the workshop on \textit{Nonequilibrium Physics from Classical to Quantum Low Dimensional Systems} where some interesting discussions took place. The work of M.K. was supported by the NSF under Grant No. DMR-0348358. F.F. acknowledges the PRIN 2007JHLPEZI.

\appendix

\section{Action for the DFP solution}
\label{ActionApp}

In this appendix, we will revise the calculation presented in \cite{abanov05} and adapt it to our case. We want to find the value of the hydrodynamic action calculated on a given solution satisfying the DFP boundary conditions.

The gradientless hydrodynamic action in imaginary time $\tau \equiv \ii t$ can, in general, be written as
\be
   {\cal S}
   = \int \de^2 x \; {\cal L} [v,\rho]
   = \int \de x \, \de \tau \, \rho \left\{ {v^2 \over 2} +
   \epsilon (\rho) -\mu \right\} \, ,
   \label{hydroact}
\ee
where
\be
   \epsilon (\rho) = {\lambda^2 \pi^2 \over 6} \rho^2
   \label{CSepsilon}
\ee
is the internal energy per particle of a Calogero system and
\be
   \mu \equiv \partial_\rho \left[ \rho \epsilon(\rho) \right]_{\rho = \rho_0} = {\lambda^2 \pi^2 \over 2} \rho_0^2
\ee
is the chemical potential.
The action (\ref{hydroact}) has to be supplemented with the continuity equation
\be
   \partial_\tau \rho + \partial_x \left( \rho v \right) = 0 \; ,
   \label{continuity}
\ee
which can be considered as a constraint relating the two conjugated fields $\rho$ and $v$. This constraint can be resolved by introducing the displacement field $\phi(x,\tau)$:
\be
   \rho = \rho_0 + \partial_x \phi \; , \qquad \qquad
   j = \rho v = - \partial_\tau \phi \; .
   \label{phidef}
\ee
Physically, the displacement field counts the number of particles to the left of a point. We can use (\ref{phidef}) to write the Lagrangian as a functional of $\phi$: ${\cal L} [\rho, v] = {\cal L} [\phi]$. Its variation then gives the Euler equation for the fluid, which can be written more simply as
\be
   \partial_\tau v + v \partial_x v = \partial_x \partial_\rho \left[ \rho \epsilon(\rho) \right] = \lambda^2 \pi^2 \rho \partial_x \rho \; .
\ee

For the particular choice of internal energy (\ref{CSepsilon}), corresponding to the Calogero-Sutherland interaction or exclusion statistics, the Euler equation and the continuity equation can be combined into a single complex Riemann-Hopf equation:
\be
   \partial_\tau k - \ii k \partial_x k = 0 \; , \qquad \qquad
   k (\tau,x) \equiv \lambda \pi \rho (\tau,x) + \ii v (\tau,x) \; .
\ee
This equation has a simple, implicit, solution of the form
\be
   k = F(x + \ii k \tau) \; .
   \label{ksolapp}
\ee

In the body of the paper, we argued that a solution satisfying the DFP boundary conditions is of the form
\be
   F(z) \equiv F(z; \rho_0, \eta) = \lambda \pi \rho_0
   + \lambda \pi \eta \left( {z \over \sqrt{z^2 - R^2}} - 1 \right) \; .
   \label{Fdef}
\ee
Here, $\rho_0$ is the background (equilibrium) density at infinity (where moreover $v=0$), and $\eta$ is a, possibly complex, constant specifying the DFP.

To calculate the Depletion Formation Probability, we need to compare the action (\ref{hydroact}) calculated on the solution (\ref{ksolapp},\ref{Fdef}) to the action of an equilibrium configuration:
\be
   {\cal S} - {\cal S}_0 = \int \de x \, \de \tau
   \left\{ \rho {v^2 \over 2} + \rho \epsilon (\rho)
   - \rho_0 \epsilon (\rho_0) - \mu (\rho - \rho_0) \right\} \, .
   \label{DFPact}
\ee
To take advantage of the fact that (\ref{ksolapp}) is a solution of the equations of motion, we first take the variation of (\ref{DFPact}) with respect to the parameters of the solution. In this way, we will reduce a two-dimensional integration to a contour integral over the boundaries, since the bulk terms are proportional to the Euler-Lagrange equations and vanish:
\be
  \de \left( {\cal S} - {\cal S}_0 \right) =
  \partial_{\rho_0} \left( {\cal S} - {\cal S}_0 \right) \; \de \rho_0
  + \partial_\eta \left( {\cal S} - {\cal S}_0 \right) \; \de \eta
  + \partial_{\bar{\eta}} \left( {\cal S} - {\cal S}_0 \right)
  \; \de \bar{\eta} \; .
  \label{diffS}
\ee

We have:
\bea
   \partial_\eta \left( {\cal S} - {\cal S}_0 \right) & = &
   \int \de^2 x \left\{
   - v \partial_\tau \phi_\eta
   - \left[ {v^2 \over 2}
   - \partial_\rho \left( \rho \epsilon \right) + \mu \right]
   \partial_x  \phi_\eta \right\}
   \nonumber \\
   & = & \int \de^2 x \left\{
   - \partial_\tau \left[ v \, \phi_\eta \right]
   - \partial_x \left[ \left( {v^2 \over 2}
   - \partial_\rho \left( \rho \epsilon \right) + \mu \right)
   \phi_\eta \right]
    \right.
   \nonumber \\
   && \qquad \quad
   + \Big[ \partial_\tau v + v \partial_x v
   - \partial_x \partial_\rho \left( \rho \epsilon \right) \Big]
   \phi_\eta \bigg\}
   \nonumber \\
   & = & - \oint \left\{
   \left[ v \, \phi_\eta \right] \de x
   + \left[ \left( {v^2 \over 2}
   - \partial_\rho \left( \rho \epsilon \right) + \mu \right)
   \phi_\eta \right] \de t \right\} \; ,
   \label{Seta}
\eea
where $\phi_\eta \equiv \partial_\eta \phi$.

The boundaries over which the contour integral is taken are, by Stoke's theorem, the points where the integrand has a discontinuity. It is easy to check that this contour comprises only two paths: one at infinity ($C_0 \equiv \{|x+ \ii k \tau| = \infty \}$) and one around the branch cut of (\ref{Fdef}), which we take along the real axis ($C_1 = \{ \tau=0^\pm, -R<x<R \}$).

From (\ref{ksolapp},\ref{Fdef}), at infinity we have
\bea
   \rho (|z_0| \to \infty) & \simeq & \rho_0
   + {R^2 \over 4} \left( {\eta \over z_0^2}
   + {\bar{\eta} \over \bar{z}_0^2} \right) + \ldots \; ,
   \label{rhoinf}
   \\
   v (|z_0| \to \infty) & \simeq &
   -\ii \lambda \pi {R^2 \over 4} \left( {\eta \over z_0^2}
   - {\bar{\eta} \over \bar{z}_0^2} \right) + \ldots \; ,
   \label{vinf}
   \\
   \phi (|z_0| \to \infty) & \simeq & - {R^2 \over 4}
   \left( {\eta \over z_0}
   + {\bar{\eta} \over \bar{z}_0} \right) + \ldots \; ,
   \label{phiinf}
\eea
and we see that the integrand in (\ref{Seta}) along $C_0$ vanishes too fast and the contour integral gives no contribution.

Close to the cut on the real axis we have
\bea
   \rho (\tau = 0^\pm; -R<x<R) & = & \rho_0 - {\eta + \bar{\eta} \over 2}
   \mp \ii {\eta - \bar{\eta} \over 2} {x \over \sqrt{R^2 - x^2}} \; ,
   \label{rhocut}
   \\
   v (\tau = 0^\pm; -R<x<R ) & = &
   \ii \lambda \pi {\eta - \bar{\eta} \over 2}
   \mp \lambda \pi {\eta + \bar{\eta} \over 2}
   {x \over \sqrt{R^2 - x^2}} \; ,
   \label{vcut}
   \\
   \phi (\tau = 0^\pm; -R<x<R) & = & - {\eta + \bar{\eta} \over 2} x
   \pm \ii {\eta - \bar{\eta} \over 2} \sqrt{ R^2 - x^2} \; .
   \label{phicut}
\eea
Therefore
\bea
   \partial_\eta \left( {\cal S} - {\cal S}_0 \right) & = &
   \int_{-R}^R \left[ v(x,0^+) \phi_\eta (x,0^+)
   - v(x,0^-) \phi_\eta (x,0^-) \right] \de x
   \nonumber \\
   & = & { \lambda \pi^2 R^2 \over 2} \bar{\eta} \; .
   \label{Setares}
\eea

Similarly, we have
\bea
   \partial_{\bar{\eta}} \left( {\cal S} - {\cal S}_0 \right) & = &
   - \oint \left\{
   \left[ v \, \phi_{\bar{\eta}} \right] \de x
   + \left[ \left( {v^2 \over 2}
   - \partial_\rho \left( \rho \epsilon \right) + \mu \right)
   \phi_{\bar{\eta}} \right] \de t \right\}
   \nonumber \\
   & = & { \lambda \pi^2 R^2 \over 2} \eta \; .
   \label{Sbaretares}
\eea

The derivative with respect to $\rho_0$ is a bit more complicated as it involves more terms. After a bit of algebra and an additional integration by parts we obtain:
\bea
   \partial_{\rho_0} \left( {\cal S} - {\cal S}_0 \right) & = &
   - \oint \bigg\{
   \Big[ v \left( \phi_{\rho_0} - x \right) \Big] \de x
   \nonumber \\
   && \quad \left.
   + \left[ \left( {v^2 \over 2}
   - \partial_\rho \left( \rho \epsilon \right) + \mu \right)
   \left( \phi_{\rho_0} + x \right)
   + \left( \partial_{\rho_0} \mu \right) \phi \right] \de t \right\}
\eea
where $\phi_{\rho_0} \equiv \partial_{\rho_0} \phi$.
Substituting the behaviors (\ref{rhoinf}-\ref{phiinf}) and (\ref{rhocut}-\ref{phicut}), the integrals around the two contours gives equal but opposite results ($\pm {1 \over 2} \lambda \pi^2 (\eta + \bar{\eta}) R^2$), which means
\be
   \partial_{\rho_0} \left( {\cal S} - {\cal S}_0 \right) = 0 \; ,
   \label{Srho0res}
\ee
as one could have expected.

We can now integrate (\ref{diffS}) using (\ref{Setares},\ref{Sbaretares},\ref{Srho0res}) to find
\be
   {\cal S}_{\rm DFP} = {\cal S} - {\cal S}_0 =
   { 1 \over 2} \, \lambda \, \pi^2 \, \eta \, \bar{\eta} \, R^2 \; .
   \label{DFPactionResult}
\ee

\section{Linearized Hydrodynamics and DFP}
\label{sec:bosonization}

Let us linearize the hydrodynamic equations, by expanding the theory and retaining only the quadratic part of the Lagrangian. This {\it linearized hydrodynamics} is usually referred in the literature on one-dimensional models as {\it bosonization}.

From the previous section, we have that the gradientless hydrodynamic Lagrangian for a one-component system is
\be
   {\cal L} [j,\rho] = {j^2 \over 2 \rho}
   + \rho \Big[ \epsilon (\rho) - \mu \Big] \; ,
\ee
where $j = \rho v$.
We expand the fields around a background value and we parametrize the fluctuations around this background through the {\it displacement field} $\phi$ as in (\ref{phidef}), so that the constraint (\ref{continuity}) is automatically satisfied. Keeping terms up to the quadratic order we have
\bea
   {\cal L} [\phi] & = &
   {1 \over 2 \rho_0} \left( \partial_\tau \phi \right)^2
   + {1 \over 2} \partial^2_\rho \left( \rho \epsilon \right)_{\rho = \rho_0}
   \left( \partial_x \phi \right)^2
   + \rho_0 \Big[ \epsilon (\rho_0)  - \mu \Big] \ldots
   \nonumber \\
   & = & {\kappa \over 2 u} \left( \partial_\tau \phi \right)^2 +
   {\kappa \, u \over 2} \left( \partial_x \phi \right)^2
   + {\cal L} [0,\rho_0] + \ldots \; ,
\eea
where in the last line we introduce the standard parameters of bosonization: the interaction parameter $\kappa=\kappa(\rho_0)$ and the sound velocity $u=u(\rho_0)$.
The displacement field evolves according to a linear wave equation:
\be
   \partial_\tau^2 \phi + u^2 \partial_x^2 \phi = 0 \; .
\ee

The linearized treatment is valid for small fluctuations around the background $\rho_0$, i.e. as long as the gradients of $\phi$ are small and only low energy excitations are involved. For this reason, it is not possible to calculate the EFP through standard bosonization, but we can consider a DFP with very small depletion.

It is simple to see \cite{abanov05} that the solution that satisfies the DFP boundary conditions is of the form:
\be
   \phi (\tau, x) = \R \left[ \eta \left( \sqrt{z_0^2 - R^2} - z_0
   \right) \right] \; ,
   \label{linsol}
\ee
where $z_0 \equiv x + \ii u(\rho_0) \tau$. For this solution to be compatible with the linearized approximation we need $|\eta|/\rho_0 \ll 1 $.

It is easy to calculate the DFP by evaluating the linearized action
\be
   {\cal S} -  {\cal S}_0 \simeq \int \de x \de \tau \left\{
   {\kappa \over  2 u} \left( \partial_\tau \phi \right)^2 +
   {\kappa \, u \over 2} \left( \partial_x \phi \right)^2 \right\}
   \label{linearaction}
\ee
on the solution (\ref{linsol}).
One immediately observes that, at zero temperature, this action does not depend on the sound velocity, as we can rescale the time as $y \equiv u \tau$:
\be
   {\cal S} -  {\cal S}_0 \simeq {\kappa \over 2} \int \de x \de y \left\{
   \left( \partial_y \phi \right)^2 +
   \left( \partial_x \phi \right)^2 \right\} \; .
   \label{rescaledS}
\ee
We can further rescale the lengths by $R$ and substituting (\ref{linsol}) in
(\ref{rescaledS}) we get
\be
   {\cal S} -  {\cal S}_0 \simeq {\kappa \over 2} \, \eta \bar{\eta} R^2 \int
   \de \tilde{x} \de \tilde{y} \left|
   \tilde{\phi}' (\tilde{x} + \ii \tilde{y}) \right|^2 \; ,
\ee where
\be
  \tilde{\phi} (z) = \sqrt{z^2 - 1} -z \; .
\ee
Now all the physical parameters have been explicitly extracted and one has just to perform an integral that contributes only with a numerical factor. The result is
\be
   {\cal S} -  {\cal S}_0 \simeq
   {\pi \over 2} \, \kappa \, \eta \, \bar{\eta} R^2 \; .
   \label{linearDFPres}
\ee
Since for a Calogero-Sutherland system $\kappa = \lambda \pi$, we notice that (\ref{linearDFPres}) exactly coincide with (\ref{DFPactionResult}). This is quite surprising, since, as we argued above, the linearized result should be trusted to be approximately correct only for small depletions. However, the Calogero kind of interaction is very special and we can extend (\ref{linearDFPres}) to higher depletion, without loosing accuracy. In section \ref{sec:discussion} we discuss the meaning of this observation.
Let us remark that this result is specific for the EFP and it is not to say that for Calogero-Sutherland systems the effects of non-linearity are in general not important. For instance, effects of non-linear spin-charge interactions were observed and discussed in \cite{KFA09}.

This DFP calculation can also be performed using the line integral technique explained in the previous section. In this case, the variation of the action (\ref{linearaction}) gives simply:
\bea
   \partial_\eta \left( {\cal S} - {\cal S}_0 \right) & \simeq &
   \oint \bigg\{
   {\kappa \over u} \left( \partial_\tau \phi \right) \phi_\eta \, \de x
   + \kappa \, u \left( \partial_x \phi \right) \phi_\eta \, \de \tau \bigg\}
   \nonumber \\
   & = & \kappa \oint \bigg\{
   \left( \partial_y \phi \right) \phi_\eta \, \de x
   + \left( \partial_x \phi \right) \phi_\eta \, \de y \bigg\} \, .
\eea
One can then proceed as we showed in the previous section to easily recover (\ref{linearDFPres}).

\end{document}